# Performance evaluation of on-chip wavelength conversion based on InP/In$_{1-x}$Ga$_x$As$_y$P$_{1-y}$ semiconductor waveguide platforms

Jin Wen[1, 5, 6*], Kang Li[2], Yongkang Gong[3], Ben Hughes[4], Michael A. Campbell[4], Mattia Lazzaini[4], Lina Duan[1], Chengju Ma[1], Wei Fan[1], Zhenan Jia[1, 5, 6], Haiwei Fu[1, 5, 6] and Nigel Copner[2*]

[1]*School of Science, Xi'an Shiyou University, Xi'an 710065, China*

[2]*Wireless and Optoelectronics Research and Innovation Centre, Faculty of Computing, Engineering and Science, University of South Wales, CF37 1DL, United Kingdom*

[3]*School of Physics and Astronomy, Cardiff University, Cardiff, CF24 3AA, United Kingdom*

[4]*National Physical Laboratory, Hampton Road, Teddington, Middlesex, TW11 0LW, United Kingdom*

[5]*Shaanxi Engineering Research Center of Oil and Gas Resource Optical Fiber Detection, Xi'an 710065, China*

[6]*Shaanxi Key Laboratory of Measurement and Control Technology for Oil and Gas wells, Xi'an 710065, China*

*\*Corresponding author: wenjin@xsyu.edu.cn, nigel.copner@southwales.ac.uk*

## Abstract

We propose and design the high confinement InP/In$_{1-x}$Ga$_x$As$_y$P$_{1-y}$ semiconductor waveguides and report the results of effective wavelength conversion based on this platform. Efficient confinement and mode field area fluctuation at different wavelength is analyzed to achieve the high nonlinear coefficient. The numerical results show that nearly zero phase-mismatch condition can be satisfied through dispersion tailoring of InP/In$_{1-x}$Ga$_x$As$_y$P$_{1-y}$ waveguides, and the wavelength conversion ranging over 40 nm with the maximum conversion efficiency -26.3 dB is achieved for fixing pump power 100 mW. Meanwhile, the influences of the doping parameter y and pumping wavelength on the bandwidth and conversion efficiency are also discussed and optimized. It is indicated the excellent optical properties of the InP/In$_{1-x}$Ga$_x$As$_y$P$_{1-y}$ waveguides and pave the way towards direct integration telecom band devices on stand semiconductor platforms.

**Key words**: InP/In$_{1-x}$Ga$_x$As$_y$P$_{1-y}$ waveguide, wavelength conversion, four-wave mixing, nonlinear optics

## 1. Introduction

Integration of III-V light emitters and amplifiers on silicon substrates is the dominated method for high-functionality and low-cost photonic integrated circuits [1, 2]. In this situation, III-V nano-lasers with extremely compact footprint and ultra-low dissipation could benefit silicon-based photonic integrated circuits in terms of integration density and power consumption [3-5]. To be compatible with passive silicon devices, the light source must emit at the transparent wavelength window in order to minimize the propagation loss. Certainly, the minimum coupling loss of fibers need light emit at the suitable wavelength bands simultaneously [6]. Therefore, many researches are supplied to replying the challenge of extending the laser spectra and reducing the loss based on III-V semiconductor platforms [7, 8]. Specially, the room-temperature InP/InGaAs nano-ridge lasers grown on silicon substrates have been demonstrated experimentally and the laser wavelength emitting can be tuned in broadband wavelength range from O band to C band with ultra-low thresholds in 2018 [9].

To fully exploit the integration of devices based on InP wafer, the active material should be designed carefully to bridge the light source and other passive devices. The InP wafer is the more appropriate platform candidate for monolithic integration of active and passive devices compared with silicon wafer because silicon is an indirect band gap semiconductor and therefore has a very low light emission efficiency [10]. Additionally, the wavelength converter based on III-V material such as InGaAsP on InP wafer can greatly extend the wavelength range through nonlinear process such as four-wave mixing [11], which can be found potential applications in various fields since the III-V ridge source can realize stable room-temperature lasing at the telecom band [12-14]. Moreover, the InGaAsP on InP wafer can supply strong on-chip mode confinement and directly integrated with the III-V laser source with low coupling loss compared with the parametric amplifiers and wavelength converters based on silicon platforms [15-17]. Recently research shows that the carrier lifetime in III-V semiconductor materials can be reduced to 0.42 ps [18], which can reduce the nonlinear loss in the telecom band and has the potential advantage for high conversion efficiency.

The semiconductor material $In_{1-x}Ga_xAs_yP_{1-y}$ has generated much interest since it can be grown on InP without lattice mismatch over the composition range $0 \leq y \leq 1$, provided $x=0.466y$ [19]. The recently research work shows the band-gap energy $E_g$ and band-gap wavelength $\lambda_g$ of the $In_{1-x}Ga_xAs_yP_{1-y}$ (y=0.8, x=0.37) matched to InP can be tuned to 0.85 eV and 1459 nm respectively, which indicates that the wavelength conversion in telecom band is feasible [20]. Moreover, the refractive index between the InP cladding and core $In_{1-x}Ga_xAs_yP_{1-y}$ can provide the relative high refractive index contrast for the better mode confinement. These features can further overcome the difficulties of on-chip integration when use the silicon waveguide on SOI wafer caused by the high nonlinear loss in telecom band and coupling loss. Therefore, the $InP/In_{1-x}Ga_xAs_yP_{1-y}$ waveguides can be used as the candidates to realize the efficient wavelength conversion with the highly integrated device.

In this work, we propose and design $InP/In_{1-x}Ga_xAs_yP_{1-y}$ waveguides for the wavelength conversion under the condition of low phase-mismatch. In the design process, $In_{1-x}Ga_xAs_yP_{1-y}$ quantum wells are separated by InP spacers and the multi-layer $InP/In_{1-x}Ga_xAs_yP_{1-y}$ waveguides can be obtained. Meanwhile, highly confinement of the waveguides are realized and optimized to reach effective mode area. The numerical results show efficient wavelength conversion based on $In_{1-x}Ga_xAs_yP_{1-y}$ waveguide achieved with the efficiency up to -26.3 dB in over 40 nm wavelength range under the phase-matching condition. The pump power used is 100 mW and the length of waveguide is just 5 mm, which realizes the relative lower power consumption and compact device on $InP/In_{1-x}Ga_xAs_yP_{1-y}$ platforms.

## 2. Design and numerical modeling

The $InP/In_{1-x}Ga_xAs_yP_{1-y}$ waveguide has the good confinement for optical field in the near telecom wavelength range. The refractive index of $In_{1-x}Ga_xAs_yP_{1-y}$ ($0 \leq y \leq 1$, $x=0.466y$) can be calculated as shown in Fig. 1 (a) [21]. It is shown the refractive index of the $In_{0.63}Ga_{0.37}As_{0.8}P_{0.2}$ is 3.58 at 1550 nm. Meanwhile, the refractive index of the InP is depict as shown in Fig. 1 (b) and it can be also found the refractive index of InP is 3.17 at 1550 nm. The refractive index increase as the value of y enhances, which is caused by the high refractive index ratio doped GaAs on the InP wafer. The scheme of the InP/In$_{1-x}$Ga$_x$As$_y$P$_{1-y}$ waveguide is shown as Fig. 2 (a). The upper cladding and lower cladding are the InP and the core is the In$_{1-x}$Ga$_x$As$_y$P$_{1-y}$, respectively. This kind of design can cause high confinement between the core and cladding due to the high refractive index contrast. From the viewpoint of fabrication, this kind of multi-layer InP/In$_{1-x}$Ga$_x$As$_y$P$_{1-y}$ structure can be easily realized [22]. The relative parameters of the waveguide are as following: W is the width of the InP/In$_{1-x}$Ga$_x$As$_y$P$_{1-y}$ waveguide and h is the height of the core In$_{1-x}$Ga$_x$As$_y$P$_{1-y}$. While h$_1$ and h$_2$ is the height of the upper cladding and lower cladding of the InP, respectively. The width and height of InP wafer substrate is 5 μm and 6 μm, respectively.

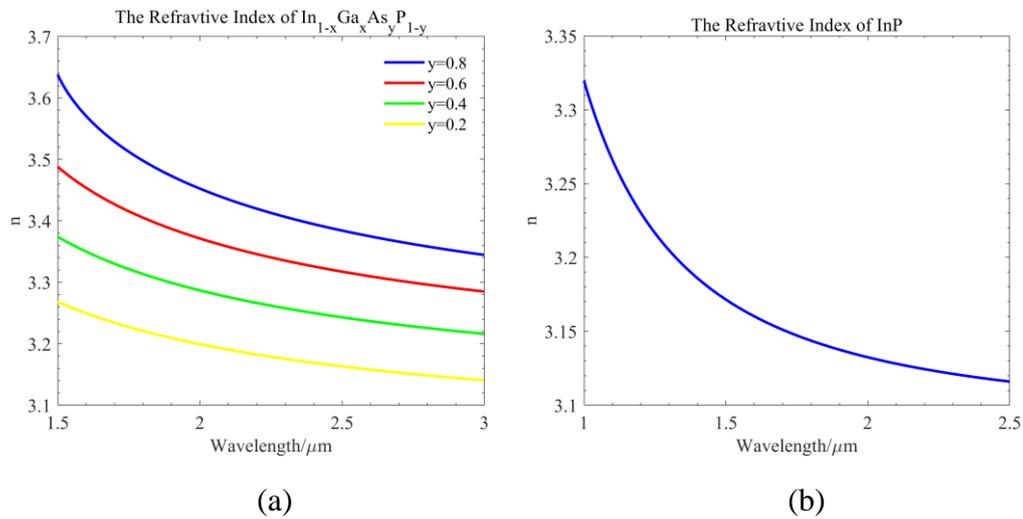

(a)                                    (b)

Fig. 1 (a) Theoretical values of refractive index n for In$_{1-x}$Ga$_x$As$_y$P$_{1-y}$ as a function of wavelength and y increments of 0.2. (b) Theoretical values of refractive index n for InP as a function of wavelength.

Next we use the finite element method (FEM) to calculate the TE mode distribution of the InP/In$_{0.63}$Ga$_{0.37}$As$_{0.8}$P$_{0.2}$ (x=0.37, y=0.8) waveguide at 1550 nm with the parameters W=2 μm, h$_1$=500 nm, h=250 nm and h$_2$=1000 nm, respectively. As shown in Fig. 2 (b), the mode distribution is focused on the core In$_{1-x}$Ga$_x$As$_y$P$_{1-y}$ part and nearly no mode distribution is leaked to the upper and lower InP cladding. Meanwhile, the calculated effective mode area is 0.73 μm$^2$, as shown the inset of Fig. 2 (b).

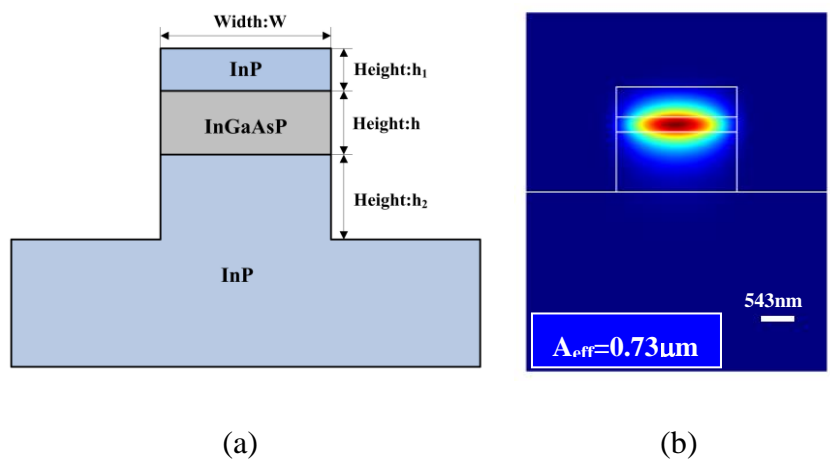

(a)          (b)

Fig. 2 (a) Scheme of InP/In$_{1-x}$Ga$_x$As$_y$P$_{1-y}$ waveguide. The parameters are as following: W is the width of the waveguide. Here h$_1$ and h$_2$ is the height of upper and lower InP cladding, respectively. h is the height of core In$_{1-x}$Ga$_x$As$_y$P$_{1-y}$ layer. (b) Calculated TE mode distribution of the InP/ In$_{1-x}$Ga$_x$As$_y$P$_{1-y}$ waveguide for the fixed geometry parameters as following: W=2 μm, h$_1$=500 nm, h=250 nm and h$_2$=1000 nm. Inset: A$_{eff}$ is the effective mode area.

In order to realize wavelength conversion based on the semiconductor platforms, the dispersion tailoring must be considered. The effective refractive index of the InP/ In$_{1-x}$Ga$_x$As$_y$P$_{1-y}$ waveguide n$_{eff}$ and the effective mode area A$_{eff}$ variation as the wavelength changing is shown in Fig. 3 based on the scheme of the waveguide as shown in Fig. 2 (a). The refractive index of the InP and $In_{1-x}Ga_xAs_yP_{1-y}$ is obtained from the results in Fig. 1 in process of calculating the effective refractive index of the $InP/In_{0.63}Ga_{0.37}As_{0.8}P_{0.2}$ waveguide. Meanwhile, when we obtain the effective mode area (for TE mode), it is aimed to realize the maximum light confinement and minimize the effective mode area $A_{eff}$ in the waveguides given by

$$A_{eff} = \frac{\left[\int_{-\infty}^{\infty}\int_{-\infty}^{\infty}|E(x,y)|^2\,dxdy\right]^2}{\int_{-\infty}^{\infty}\int_{-\infty}^{\infty}|E(x,y)|^4\,dxdy} \quad (1)$$

Here E(x, y) is the electric field of the mode [23].

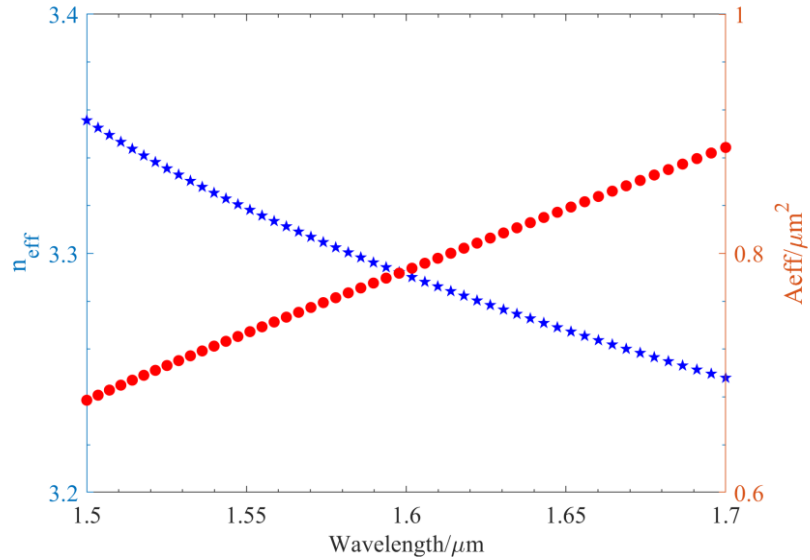

Fig. 3 Calculated features of the $InP/In_{0.63}Ga_{0.37}As_{0.8}P_{0.2}$ waveguide as a function of wavelength. Red circle line: effective refractive index as the wavelength changing. Blue pentagon line: effective mode area as the wavelength changing (TE mode). The parameters of the waveguide are the same as show in Fig. 2.

There are multi-parameters used to optimized for low phase-mismatch and near zero-dispersion condition. Here the parameters used in the simulation is W=2 μm, $h_1$=500 nm and $h_2$=1000 nm. The height of $In_{0.63}Ga_{0.37}As_{0.8}P_{0.2}$ changes from 200 nm to 750 nm. The effective refractive index $n_{eff}$, the first dispersion coefficient $\beta_1$, the second order dispersion coefficient $\beta_2$ and the fourth order dispersion coefficient $\beta_4$ are plotted in Fig. 4 when the parameter h changes from 200 nm to 500 nm. It is remarkably seen the second order dispersion coefficient decreased as the h increases. However, the zero-dispersion wavelengths can't appear in the wavelength range between 1500 nm to 1700 nm, which is clearly shown in Fig. 3 (c). Meanwhile, the zero-dispersion wavelengths can be tuned to around 1550 nm under the condition of optimizing the parameters h in the range of 600 nm to 750 nm. Additionally, the height of upper InP cladding $h_1$ has also the impact on the dispersion adjusting. The first order dispersion coefficient $\beta_1$, the dispersion, the second order dispersion $\beta_2$ and the fourth order dispersion coefficient $\beta_4$ can be found in Fig. 5 when the height of core $In_{0.63}Ga_{0.37}As_{0.8}P_{0.2}$ is fixed at 700 nm. It is clearly shown that when the height of upper InP cladding $h_1$ changes from 200 nm to 500 nm, the flat dispersion curve changes to be steep around 1550 nm region. Meanwhile, the zero-dispersion wavelength shifts to longer wavelength range. It is also found the variation of high order dispersion coefficients is small when $h_1$ changes from 400 nm to 500 nm, which means the optimized upper InP cladding $h_1$ is in this range. The numerical results indicate that impact of the $h_1$ on the dispersion tailoring is weak than that of h. The crossover points of the fourth order dispersion coefficients are different, which also demonstrates different dispersion slope for changing the parameter h and $h_1$.

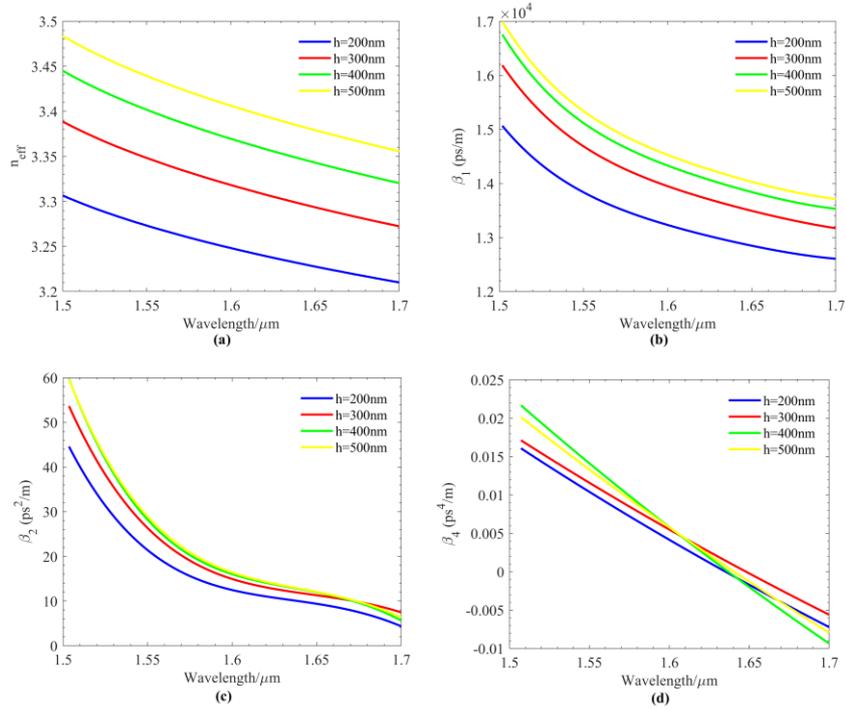

Fig. 4 High order dispersions of InP/In$_{0.63}$Ga$_{0.37}$As$_{0.8}$P$_{0.2}$ waveguides as function of wavelength for different value of h. (a) Effective refractive index n$_{eff}$ (b) First order dispersion coefficient $\beta_1$ (c) Second order dispersion coefficient $\beta_2$ and (d) Fourth order dispersion coefficient $\beta_4$.

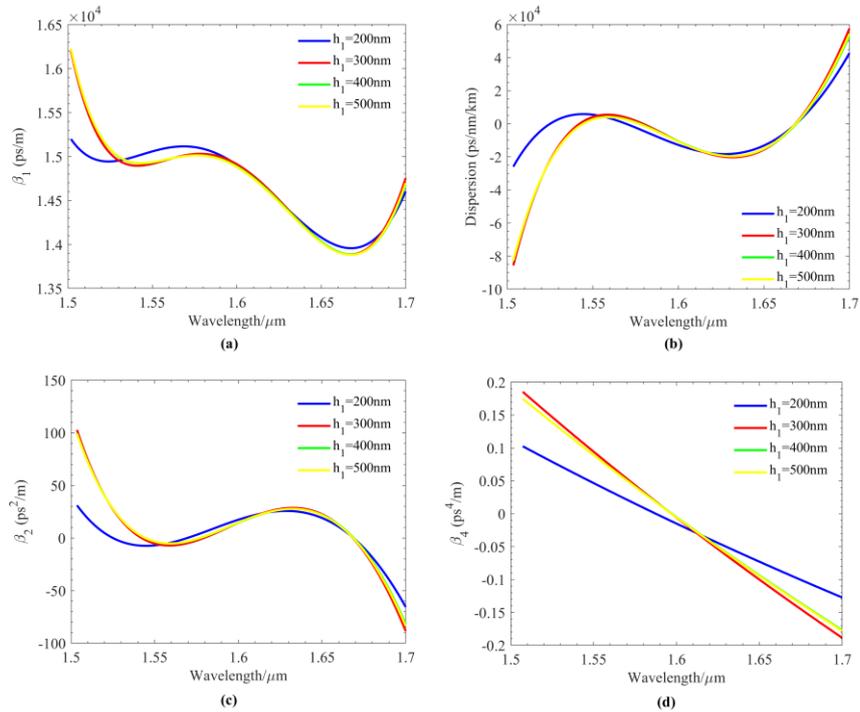

Fig. 5 High order dispersions of InP/In$_{0.63}$Ga$_{0.37}$As$_{0.8}$P$_{0.2}$ waveguides as function of wavelength for different value of h$_1$. (a) First order dispersion coefficient β$_1$ (b) Dispersion (c) Second order dispersion coefficient β$_2$ and (d) Fourth order dispersion coefficient β$_4$.

Furthermore, we give the numerical model of wavelength conversion based on InP/In$_{1-x}$Ga$_x$As$_y$P$_{1-y}$ (0≤y≤1, x=0.466y) waveguides. The theoretical basis of the wavelength conversion is four-wave mixing and the numerical model of the wavelength conversion based on the InP/In$_{1-x}$Ga$_x$As$_y$P$_{1-y}$ waveguide platform can be described by the nonlinear coupled equations as following in [24]:

$$\frac{dA_p}{dz} = -\frac{1}{2}[\alpha + \frac{\beta_{TPA}}{A_{eff}}(|A_p|^2 + 2|A_s|^2 + 2|A_i|^2)]A_p + i\gamma_p(|A_p|^2 + 2|A_s|^2 + 2|A_i|^2)A_p + 2i\gamma_p A_s A_i A_p^* \exp(i\Delta kz) \quad (2)$$

$$\frac{dA_s}{dz} = -\frac{1}{2}[\alpha + \frac{\beta_{TPA}}{A_{eff}}(|A_s|^2 + 2|A_p|^2 + 2|A_i|^2)]A_s + i\gamma_s(|A_s|^2 + 2|A_p|^2 + 2|A_i|^2)A_s + 2i\gamma_s A_p^2 A_i^* \exp(-i\Delta kz) \quad (3)$$

$$\frac{dA_i}{dz} = -\frac{1}{2}[\alpha + \frac{\beta_{TPA}}{A_{eff}}(|A_i|^2 + 2|A_p|^2 + 2|A_s|^2)]A_i + i\gamma_i(|A_i|^2 + 2|A_p|^2 + 2|A_s|^2)A_i + 2i\gamma_i A_p^2 A_s^* \exp(-i\Delta kz) \quad (4)$$

In the Eqs. (2)-(4), A$_p$, A$_s$ and A$_i$ are the amplitude of the pump, signal and idler waves respectively and z is the propagation distance along the waveguide. Here α is the linear loss coefficient of the InP/In$_{1-x}$Ga$_x$As$_y$P$_{1-y}$ waveguide. The second item of the left equation is the two-photon absorption and β$_{TPA}$ is the two-photon absorption coefficient and A$_{eff}$ is the effective mode area. The parameter γ$_j$ (j=p, s, i) is the nonlinear coefficient for pump, signal and idler wavelengths. Considering the effects including the self-phase modulation and cross-phase modulation, the phase mismatch Δk is given by [25],

$$\Delta k = 2\gamma_p P - \Delta k_{linear} \quad (5)$$

where $\gamma_p = 2\pi n_2/\lambda A_{eff}$, $n_2$ is the nonlinear refractive index of the waveguide and $A_{eff}$ is the effective mode area at the wavelength $\lambda$. P is the pump power, $\Delta k_{linear} = 2k_p - k_s - k_i$ is the linear phase-mismatch, and $k_p$, $k_s$ and $k_i$ are the pump, signal and idler propagation constants. Considering the dispersion effects up to fourth order, the linear phase-mismatch is given by,

$$\Delta k_{linear} = -\beta_2 \Omega^2 - \frac{1}{12}\Omega^4 \qquad (6)$$

where $\beta_2$ and $\beta_4$ are the second order dispersion coefficient and the fourth order dispersion coefficient as the description in the part of dispersion tailoring, and $\Omega$ is the frequency difference between the pump and signal waves.

Through solving the nonlinear coupled equations, the conversion efficiency and bandwidth can be obtained. Here we focus the flatness and efficiency of the wavelength conversion. The conversion efficiency CE can be defined as,

$$CE = \frac{P_i^{out}}{P_s^{in}} \qquad (7)$$

where $P_i^{out}$ is the output idler power and $P_s^{in}$ is the input signal power, respectively. In the process of four-wave mixing, the pump power $P_p$ transferred to the signal and idler waves, respectively. Achieving the maximum bandwidth and highest conversion efficiency, the InP/In$_{1-x}$Ga$_x$As$_y$P$_{1-y}$ waveguide must be optimized. Due to the even order dispersions play an important role in the phase-mismatch, the dispersion tailoring study is necessary.

## 3. Results and discussion

Given the dispersion tailoring and numerical model, the simulation results of wavelength conversion based on InP/In$_{1-x}$Ga$_x$As$_y$P$_{1-y}$ waveguides can be obtained. It is pointed that pump power is 100 mW and the length of waveguide is 5 mm. The linear loss coefficient α=0.5 dB/cm, the two-photon absorption coefficient $\beta_{TPA}=1\times10^{-12}$ m/W, which dominates the nonlinear loss. The nonlinear refractive index of InP/In$_{1-x}$Ga$_x$As$_y$P$_{1-y}$ waveguide given in this simulation is n$_2$=2.2×10$^{-17}$ m$^2$/W [26]. To realize the broadband conversion efficiency, the phase-mismatch is the dominated role in the process of four-wave mixing. We firstly give the numerical results of phase-mismatch under the condition of pump at wavelength 1550 nm.

Considering the high-order dispersion coefficients up to 4$^{th}$, the phase-mismatch curves can be shown as Fig. 6 for giving different parameters of h. When the height of core In$_{0.63}$Ga$_{0.37}$As$_{0.8}$P$_{0.2}$ changes from 200 nm to 750 nm with the height of upper and lower InP cladding fixed at 500 nm and 1000 nm, respectively. The near zero phase mismatch is only realized at 1550 nm. In the wavelength range of 1.5 μm, the phase-mismatch curves are almost coincided. The value of phase mismatch is over 1000/cm at 1.5 μm zone. When h changes from 600 nm to 750 nm, the phase mismatch condition varies markedly as shown in Fig. 6 (b). The changing point of h is at around 700 nm. In this process, the quasi-phase matching condition can be satisfied in the broadband wavelength range.

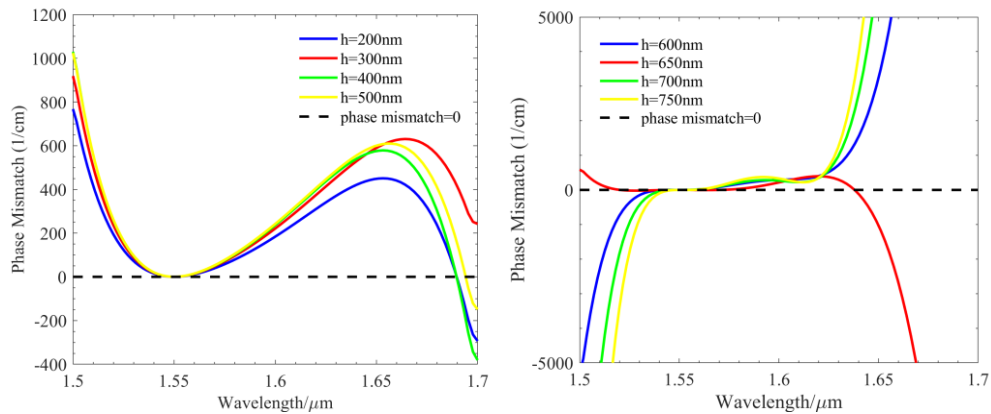

(a) (b)

Fig. 6 Phase-mismatch curves of the InP/In$_{0.63}$Ga$_{0.37}$As$_{0.8}$P$_{0.2}$ waveguides. (a) phase-mismatch of the InP/In$_{0.63}$Ga$_{0.37}$As$_{0.8}$P$_{0.2}$ waveguides when the value of h changes from 200 nm to 500 nm. (b). phase-mismatch of the InP/In$_{0.63}$Ga$_{0.37}$As$_{0.8}$P$_{0.2}$ waveguides when the value of h changes from 600 nm to 750 nm. Pump wavelength is 1550 nm.

It can be seen obviously from Fig. 6 when the height of core In$_{0.63}$Ga$_{0.37}$As$_{0.8}$P$_{0.2}$ h is changed from 600 nm to 750 nm, the near zero phase-mismatch can be obtained from about 1.5 μm to 1.6 μm, which means wavelength conversion can be realized under this phase-match condition. The numerical results of wavelength conversion are obtained under the phase-mismatch condition as shown in Fig. 7. In this simulation process, the pump power is 100 mW, the input signal and idle power are 10 mW and 0, respectively. The pumping wavelength is fixed at 1550 nm and the length of waveguide is 5 mm.

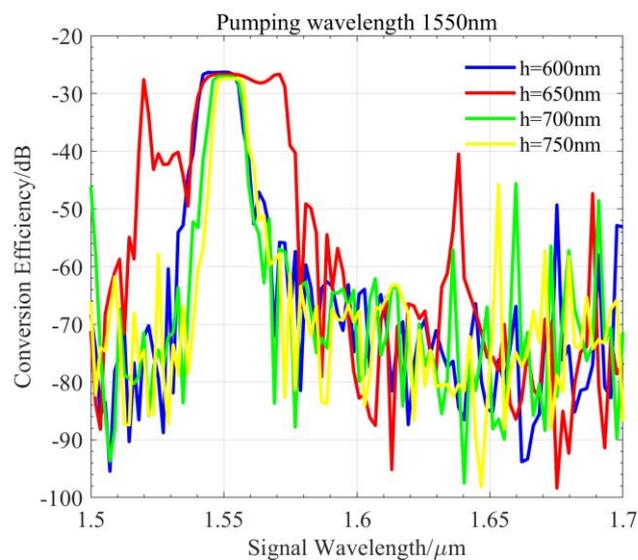

Fig. 7 Conversion efficiency of InP/In$_{0.63}$Ga$_{0.37}$As$_{0.8}$P$_{0.2}$ waveguides as function of wavelength. The height of $In_{0.63}Ga_{0.37}As_{0.8}P_{0.2}$ h changes from 600 nm to 750 nm.

As seen in Fig. 7, the conversion efficiency can reach to over -30 dB when the pump power is 100 mW. The numerical results agree with the value of phase-mismatch depicted in Fig. 6. Relative flat wavelength conversion is shown in Fig. 7 when the parameter h is optimized in the range of 600 nm to 750 nm. The conversion bandwidth is over 40 nm with the conversion efficiency over -30 dB when h is 650 nm. Moreover, the conversion efficiency will decrease when h further increases over 700 nm, especially in the longer wavelength region. It is demonstrates that the optimized height of $In_{0.63}Ga_{0.37}As_{0.8}P_{0.2}$ is in the range of 650 nm-700 nm.

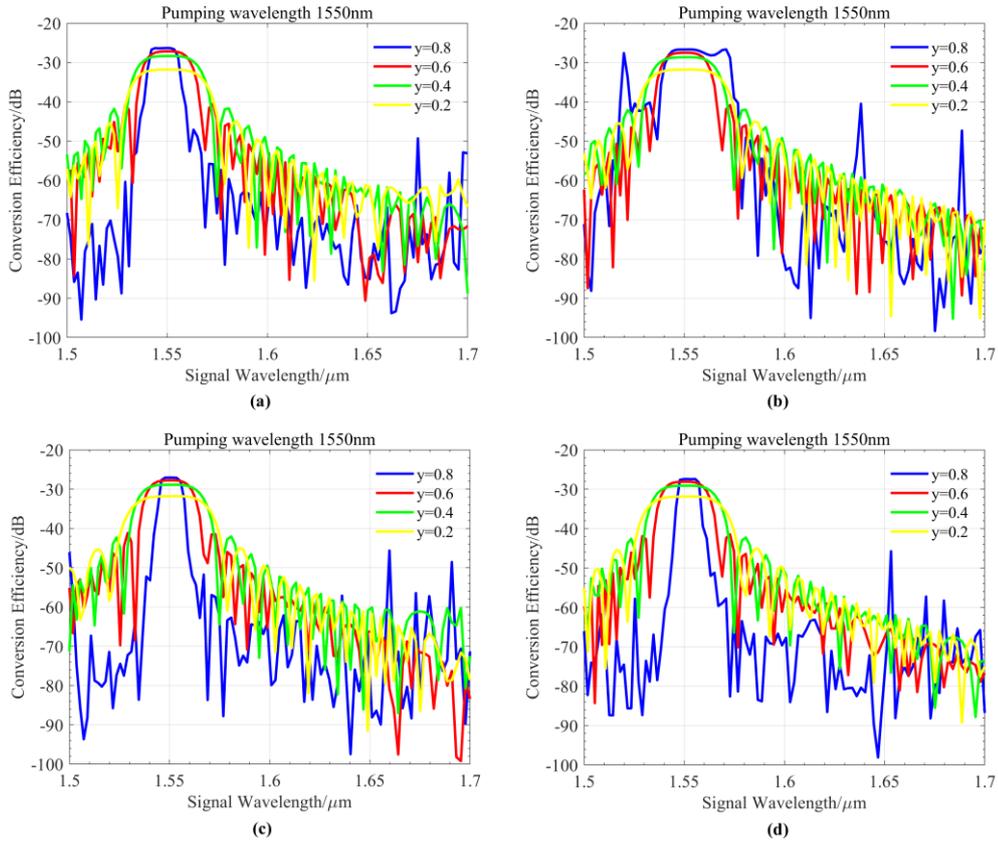

Fig. 8 Conversion efficiency of $InP/In_{1-x}Ga_xAs_yP_{1-y}$ waveguides with different parameters y and h. (a) The height of $In_{1-x}Ga_xAs_yP_{1-y}$ is 600 nm. (b) The height of

$In_{1-x}Ga_xAs_yP_{1-y}$ is 650 nm. (c) The height of $In_{1-x}Ga_xAs_yP_{1-y}$ is 700 nm. (d) The height of $In_{1-x}Ga_xAs_yP_{1-y}$ is 750 nm. The value of y changes from 0.2 to 0.8. The pumping wavelength is fixed at 1550 nm.

We will further explore the conversion efficiency based on $In_{1-x}Ga_xAs_yP_{1-y}$ waveguides through tailoring the parameters of x and y (x=0.466y). The relative refractive index of $In_{1-x}Ga_xAs_yP_{1-y}$ can be found in Fig. 1. It can be found that when the height of $In_{1-x}Ga_xAs_yP_{1-y}$ changes from 600 nm to 750 nm, the conversion efficiency decreased as the nonlinear coefficient reduced caused by the effective mode area enhances. For the fixed parameter h, the bandwidth of wavelength conversion increases, which can be obviously compared for y=0.8 and y=0.2 in Fig. 8. The maximum of conversion efficiency is -26.3 dB under the condition of pump power 100 mW. When the value of doped parameters change from 0.8 to 0.2, the refractive index of $In_{1-x}Ga_xAs_yP_{1-y}$ reduces as the same trend of the effective index of InP/ $In_{1-x}Ga_xAs_yP_{1-y}$ waveguide. The light confinement will reduce as the value of y decrease to nearly the same as the upper and lower cladding InP (as shown in Fig. 1 (b)), which gives rise to the enhance value of $A_{eff}$. That's the reason for the reduced conversion efficiency as the value of y decreases. On the other hand, the reduced effective refractive index of InP/ $In_{1-x}Ga_xAs_yP_{1-y}$ waveguide causes the lower effective refractive index contrast $\Delta n_{eff}$ between the 1.5 μm and 1.7 μm wavelength range, which changes the dispersion of the waveguide. Under this condition, broadband wavelength range can be satisfied the phase-match condition and the bandwidth enhance due to increasing value of doped parameters y.

To explore the influence of the pumping wavelength on the conversion efficiency and bandwidth, we change the pumping wavelength from 1551 nm to 1559 nm with increment 2 nm. The numerical results of wavelength conversion are shown in Fig. 9. In this simulation, the parameters of InP/ $In_{1-x}Ga_xAs_yP_{1-y}$ (y=0.8, x=0.466y) waveguide are fixed as following: the width of waveguide W=2 μm, the height of upper InP $h_1$=500 nm, the height of lower InP $h_2$=1000 nm and the height of $In_{1-x}Ga_xAs_yP_{1-y}$ is h=670 nm. It can be found that the maximum conversion efficiency is about -26.9 dB, which keeps the same value whatever the pumping wavelength changes. Meanwhile, a peak conversion efficiency can be found at about 1640 nm caused by the quasi-phase matching condition. Compared with the numerical results between the pumping wavelength 1551 nm and 1559 nm, it is shown the bandwidth of wavelength conversion enhances when the pumping wavelength extends to longer wavelength range with more flatness significantly. This phenomenon can be explained the near zero phase-matching condition can be satisfied in broadband wavelength range through adjusting the pumping wavelength just 8 nm.

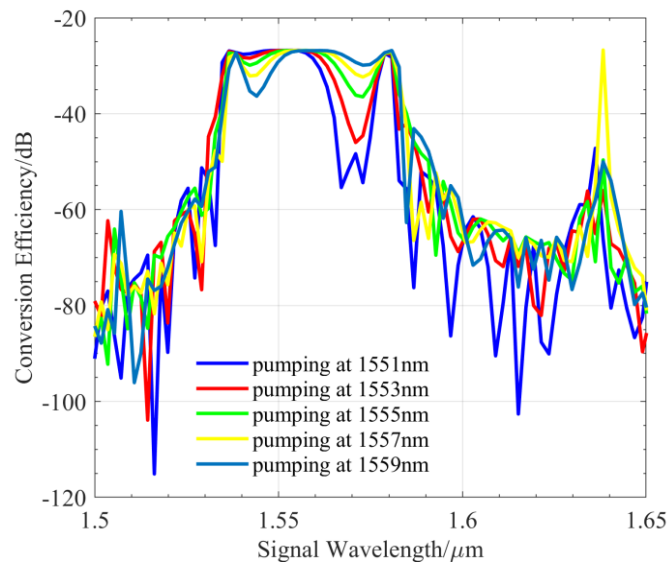

Fig. 9 Wavelength conversion based on the InP/In$_{1-x}$Ga$_x$As$_y$P$_{1-y}$ (y=0.8, x=0.466y) waveguides as the function of pumping wavelength. The pumping wavelength changes from 1551 nm to 1559 nm. The parameters of the InP/ In$_{1-x}$Ga$_x$As$_y$P$_{1-y}$ waveguide are as following: W=2 μm, h$_1$=500 nm, h=670 nm and h$_2$=1000 nm.

## 4. Conclusions

We investigate the InP/In$_{1-x}$Ga$_x$As$_y$P$_{1-y}$ waveguide platforms to realize wavelength conversion with the efficiency up to -26.3 dB and bandwidth over 40 nm. Dispersion tailoring and phase-mismatch of the InP/In$_{1-x}$Ga$_x$As$_y$P$_{1-y}$ waveguides are discussed to reveal the impacts on the conversion efficiency, bandwidth and flatness. The numerical results show that the optimized phase-mismatch can be reduced in the broadband wavelength range under the condition of fixed pump power and length of waveguide. As the value of doped parameter y in InP/In$_{1-x}$Ga$_x$As$_y$P$_{1-y}$ waveguides decrease, the bandwidth of wavelength conversion increases remarkably due to the flexible refractive index distribution in the process of changing the doped parameter y. Meanwhile, it is found that the relative flat wavelength conversion can be obtained through adjusting pumping wavelength in 8 nm (from 1551 nm to 1559 nm). This research can be applied in design and optimization of on-chip nonlinear optics.

## Acknowledgments

This work was supported by the National Natural Science Foundation of China under Grant No.61505160, the Innovation Capability Support Program of Shaanxi (Program No. 2018KJXX-042) and the Natural Science Basic Research Program of Shaanxi (Program No.2019JM-084).

# References

[1]. Z. Zhou, B. Yin, and J. Michel, "On-chip light sources for silicon photonics," Light Sci. Appl. **4**, e358 (2015).

[2]. D. Liang and J. Bowers, "Recent progress in lasers on silicon," Nature Photonics **4**, 511-517 (2010).

[3]. R. Chen, T. Tran, K. Ng, W. Ko, L. Chuang, F. Sedgwick, and C. Chang-Hasnain, "Nanolasers grown on silicon," Nature Photonics **5**, 170-175 (2011).

[4]. B. Mayer, L. Janker, B. Loitsch, J. Treu, T. Kostenbader, S. Lichtmannecker, T. Reichert, S. Morkotter, M. Kaniber, G. Abstreiter, and C. Gies, "Monolithically integrated high-β nanowire lasers on silicon," Nano Lett. **16**, 152-156 (2015).

[5]. J. Tatebayashi, S. Kako, J. Ho, Y. Ota, S. Iwamoto, and Y. Arakawa, "Room-temperature lasing in a single nanowire with quantum dots," Nature Photonics **9**, 501-505 (2015).

[6]. D. Thomson, A. Zilkie, J. Bowers, T. Komljenovic, G. Reed, L. Vivien, D. Marris-Morini, E. Cassan, L. Virot, J. Fédéli, J. Hartmann, J. Schmid, D. Xu, F. Boeuf, P. O'Brien, G. Mashanovich, and M. Nedeljkovic, "Roadmap on silicon photonics," J. Opt. **18**, 073003 (2016).

[7]. M. Takiguchi, A. Yokoo, K. Nozaki, M. Birowosuto, K. Tateno, G. Zhang, E. Kuramochi, A. Shinya, and M. Notomi, "Continuous-wave operation and 10-Gb/s direct modulation of InAsP/InP sub-wavelength nanowire laser on silicon photonic crystal," APL Photon. **2**, 046106 (2017).

[8]. H. Kim, W. Lee, A. Farrell, A. Balgarkashi, and D. Huffaker, "Telecom-wavelength bottom-up nanobeam lasers on silicon-on-insulator," Nano Lett. **17**, 5244-5250 (2017).

[9]. Y. Han, W. Ng, C. Ma, Q. Li, S. Zhu, C. Chan, K. Ng, S. Lennon, R. Taylor, K. Wong, and K. Lau, "Room-temperature InP/InGaAs nano-ridge lasers grown on Si and emitting at telecom bands," Optica **5**, 918-923 (2018).

[10]. H. Rong, R. Jones, A. Liu , O. Cohen, D. Hak, A. Fang, and M. Paniccia, "A continuous-wave Raman silicon laser," Nature **433**, 725-728 (2005).

[11]. P. Apiratikul, J. Wathen, G. Porkolab, B. Wang, L. He, T. Murphy, and C. Richardson, "Enhanced continuous-wave four-wave mixing efficiency in nonlinear AlGaAs waveguides," Optics Express **22**(22), 26814-26824 (2014).

[12]. S. Li, X. Zhou, M. Li, X. Kong, J. Mi, M. Wang, W. Wang, and J. Pan, "Ridge InGaAs/InP multi-quantum-well selective growth in nanoscale trenches on Si (001) substrate," Appl. Phys. Lett. **108**, 021902 (2016).

[13]. Y. Han, Q. Li, S. Chang, W. Hsu, and K. Lau, "Growing InGaAs quasi-quantum wires inside semi-rhombic shaped planar InP nanowires on exact (001) silicon," Appl. Phys. Lett. **108**, 242105 (2016).

[14]. Y. Han, Q. Li, S. Zhu, K. Ng, and K. Lau, "Continuous-wave lasing from InP/InGaAs nanoridges at telecommunication wavelengths," Appl. Phys. Lett. **111**, 212101 (2017).

[15]. X. Liu, R. O Jr, Y. Vlasov, W. Green, "Mid-infrared optical parametric amplifier using silicon nanophotonic waveguides," Nature Photonics **4**, 557-560 (2010)

[16]. E. Tien, Y. Huang, S. Gao, Q. Song, F. Qian, S. Kalyoncu, and O Boyraz, "Discrete parametric band conversion in silicon for mid-infrared applications," Optics Express **18**(21), 21981-21989 (2010).

[17]. J. Wen, H. Liu, N. Huang, Q. Sun, and W. Zhao, "Widely tunable femtosecond optical parametric oscillator based on silicon-on-insulator waveguides," Optics Express **20**(4), 3490-3498 (2012).

[18]. D. Cui, S. M. Hubbard, D. Pavlidis, A. Eisenbach, and C. Chelli, "Impact of doping and MOCVD conditions on minority carrier lifetime of zinc-and carbon-doped InGaAs and its applications to zinc-and carbon-doped InP/InGaAs heterostructure bipolar transistors," Semicond. Sci. Technol. **17**, 503-509 (2002).

[19]. B. Jensen and A. Torabi, "Refractive index of quaternary $In_{1-x}Ga_xAs_yP_{1-y}$ lattice matched to InP," J. Appl. Phys. **54**, 3623 (1983).

[20]. S. Saeidi, K. M. Awan, L. Sirbu, and K. Dolgaleva, "Nonlinear photonics on-a-chip in III-V semiconductors: quest for promising material candidates," Applied Optics **56**(19), 5532-5541 (2017).

[21]. B. Jensen, A. Torabi, "Quantum theory of the dispersion of the refractive index near the fundamental absorption edge in compound semiconductors," IEEE Journal of Quantum Electronics **19**(3), 448-457 (1983).

[22]. Y. Han, Q. Li, and K. M. Lau, "Highly ordered horizontal indium gallium arsenide/indium phosphide multi-quantum-well in wire structure on (001) silicon substrates," J. Appl. Phys. **120**, 245701 (2016).

[23]. G. P. Agrawal, *Nonlinear Fiber Optics*, 3rd ed. (Academic, San Diego, 2001).

[24]. K. Dolgaleva, W. Ng, L. Qian, and J. Aitchison, "Compact highly-nonlinear AlGaAs waveguides for efficient wavelength conversion," Optics Express **19**(13), 12440-12455 (2011).

[25]. M. Foster, A. Turner, R. Salem, M. Lipson, and A. Gaeta, "Broad-band continuous-wave parametric wavelength conversion in silicon nanowaveguides," Optics Express **15**(20), 12949-12958 (2007).

[26]. P. Yupapin, B. Vanishkorn, "Mathematical simulation of light pulse propagating within a microring resonator system and applications," Applied Mathematical Modelling **35**(4), 1729-1738 (2011).